\documentclass[preprintnumbers,superscriptaddress,showkeys,byrevtex]{revtex4}

\usepackage{amsmath,amsfonts,amssymb,amscd,amsxtra,amsthm}
\usepackage{graphicx}
\usepackage{epstopdf}
\usepackage{bm}
\usepackage{feynmp}
\usepackage{feynmp-auto}
\usepackage[colorlinks=true, pdfstartview=FitV, bookmarks=true, bookmarksnumbered=true, breaklinks]{hyperref}
\usepackage{mathtools,braket}
\usepackage{soul}
\usepackage{lipsum} 
\usepackage{slashed}
\usepackage{xcolor}
\usepackage{physics}
\usepackage{multirow}

\usepackage{newtxmath}
\usepackage{mathrsfs}

\definecolor{blue}{rgb}{0.0, 0.0, 1.0}
\definecolor{red}{rgb}{1.0, 0.0, 0.0}
\definecolor{royalblue}{rgb}{0.0, 0.14, 0.4}
\hypersetup{linkcolor=royalblue, citecolor=blue, urlcolor=royalblue}

\usepackage{hyperref}
\hypersetup{colorlinks=true,citecolor=blue,linkcolor=blue,urlcolor=blue}

\usepackage[mathlines]{lineno}
\usepackage{tikz,xcolor}
\definecolor{lime}{HTML}{A6CE39}
\DeclareRobustCommand{\orcidicon}{%
	\begin{tikzpicture}
	\draw[lime, fill=lime] (0,0) 
	circle [radius=0.16] 
	node[white] {{\fontfamily{qag}\selectfont \tiny ID}};
	\draw[white, fill=white] (-0.0625,0.095) 
	circle [radius=0.007];
	\end{tikzpicture}
	\hspace{-2mm}
}

\foreach \x in {A, ..., Z}{%
	\expandafter\xdef\csname orcid\x\endcsname{\noexpand\href{https://orcid.org/\csname orcidauthor\x\endcsname}{\noexpand\orcidicon}}
}

\begin{document}
\preprint{PKNU-NuHaTh-2024}
\title{Nonstandard interactions of neutrinos with dense matter}
\author{Parada T. P. Hutauruk\orcidB{}}
\email{phutauruk@pknu.ac.kr}
\affiliation{Department of Physics, Pukyong National University (PKNU), Busan 48513, Korea}
\date{\today}

\begin{abstract}
Nonstandard interaction is expected to be a crucial hint in explaining the experimental data on neutrino scattering off electrons. In this context, the nonstandard interaction vector and axial-vector couplings are needed to be extracted from recent experiments and a few of them are now available in the literature. With these coupling bounds, in this paper, I explore their impacts on the neutrinos interacting with the free electron gas in dense matter. To this end, I compute and predict the neutrino differential cross section and mean free path in dense matter for those existing experimental bounds. Interesting behavior in the neutrino cross sections and mean free path is found for the different nonstandard interaction couplings from different experiment constraints. I also found that the neutrino cross-section and mean free path in the dense matter are very sensitive to values and signs of the nonstandard interaction couplings, leading to different prediction results to the Standard Model cross-section and mean free path as well as their totals, which is given by a sum of the Standard Model and nonstandard interaction.
\end{abstract}
\keywords{nonstandard interaction, neutral current, mean free path, differential cross section, axial and axial-vector couplings}
\maketitle

\section{Introduction}
Neutrino elastic scattering plays a crucial role in the searches for neutrino properties such as neutrino mass via neutrino oscillations~\cite{Kamiokande-II:1990lll} and in the study of solar neutrino~\cite{Friedland:2004pp,Barger:1991ae}. In the neutrino oscillation cases, it was expected that solar neutrino oscillation particularly dominated by matter effects~\cite{Wolfenstein:1977ue,Wolfenstein:1979ni,Kuo:1989qe,Bahcall:1989hk}. On the other hand, neutrino interaction with matter was expected to play an important role in astrophysics and stellar structure and evolution such as the cooling process of neutron stars~\cite{Pethick:1991mk, Yakovlev:2004iq} and neutrino emission of supernova formation~\cite{Janka:2017vlw}. Neutrino interaction with matter elastically might happen either in the weak charge current (CC) or neutral current (NC) reaction processes~\cite{Kalempouw-Williams:2005zbp,Horowitz:1990it,Reddy:1998hb,Hutauruk:2023nqc,Hutauruk:2020mhl,Hutauruk:2021cgi}. In the past, it was known that neutrinos were massless particles with zero electric charge as predicted in the Standard Model (SM). Still, the recent evidence indicates that neutrinos are expected to have mass~\cite{Singh:2023ked,KATRIN:2021uub,Petcov:2013poa,Jimenez:2022dkn,Hutauruk:2020xtk}, and spatial structure~\cite{Kalempouw-Williams:2005zbp,Hutauruk:2020mhl}, which could not be accommodated and explained in the SM. The latter can only be explained in the so-called beyond the Standard Model (BSM)~\cite{Valle:1997pb,Valle:1989xn}. Several BSM scenarios have been proposed such as the neutrino electromagnetic properties~\cite{Vogel:1989iv,Raffelt:1999gv,Pulido:1991fb,Broggini:2012df,Giunti:2014ixa,Fujikawa:1980yx,Shrock:1982sc}, the new interactions mediated by light bosons~\cite{Boehm:2004uq,Langacker:2008yv}, and nonstandard neutrino interaction (NSI) with matter~\cite{Das:2017iuj,Raffelt:1992pi}. In this work, I will focus on the NC neutrino NSI with free electron gas in dense matter, which is an extension of previous work in the SM~\cite{Kalempouw-Williams:2005zbp}. Studies on the neutrino SM interaction with dense matter have been made in the previous works~\cite{Kalempouw-Williams:2005zbp,Horowitz:1990it,Reddy:1998hb,Hutauruk:2021cgi}.

The extension works to BSM have also been implemented in the neutrino interaction with matter by taking the neutrino electromagnetic form factors i.e. the neutrino charge radius and neutrino magnetic moment into account~\cite{Hutauruk:2020mhl,AtzoriCorona:2024rtv}. Authors of Ref.~\cite{Hutauruk:2020mhl} reported that the neutrino moment magnetic contributes less significantly to the differential cross-section (NDCS) and mean free path (NMFP) of the neutrino interaction with standard matter constituents in neutron stars~\cite{Hutauruk:2020mhl}, while the charge radius of neutrino has a significant contribution to the NDCS and NMFP. Nowadays such neutrino electromagnetic form factor studies become more interesting because of the strong evidence indicated by several experiments~\cite{MUNU:2003peb,Raffelt:1999gv,CHARM-II:1994aeb,CONUS:2022qbb}. Even though the existing experimental data on neutrino magnetic moment and charge radius still have quite large uncertainties. More measurements on neutrino magnetic moment and charge radius are still ongoing, and many more will be measured in the future. They will provide more precise and accurate experimental data of the neutrino magnetic moment and charge radius with less uncertainty shortly. 

Besides BSM of the neutrino electromagnetic form factors as explained above, recently the NSI of neutrino scattering off electrons attracted more attention of the neutrino physics community~\cite{Agarwalla:2012wf,Coloma:2022umy,Bolanos:2008km,Davidson:2003ha,LSND:2001akn,TEXONO:2010tnr,Escrihuela:2011cf,Ohlsson:2012kf,Berezhiani:1987gf,Super-Kamiokande:2011dam,Liao:2021rjw,Barranco:2005ps,Berezhiani:2001rs,Berezhiani:2001rt,Berezhiani:1994hy}. This is because the NSI is expected to be an important hint to describe the data and the extraction of the NSI vector and axial-vector couplings is urgently needed. Additionally, it was anticipated that the NSI would be important for the neutrino oscillation in matter, where the neutrino oscillation was dominated by matter~\cite{Wolfenstein:1977ue,Wolfenstein:1979ni} as previously explained. However, the main obstacle remaining is that the vector and axial-vector couplings of the NSI interaction with matter are still unknown. Nowadays some of the NSI vector and axial-vector coupling constraints with a few percent levels of uncertainties (at 90\% confidence level) are available in the literature. The couplings still have rather large uncertainties but in better constraints. Indeed, more work and analyses are deserved and required to extract these NSI couplings with better precision and accuracy from the cross-section of $\nu_\alpha-$e scattering experiments. With these available NSI couplings, it would be interesting to see their implication in the neutrino NSI with free electron gas in dense matter. This motivates me to compute the NDCS and NMFP to test those NSI couplings in the neutrino scattering off the free electron gas in the present work.

In this paper, I explore the implications of the experimental bounds of the NSI vector and axial-vector couplings constrained by the current solar neutrino, atmospheric neutrino, reactor, and long-baseline neutrino experiments such as Borexino Phase I~\cite{Agarwalla:2012wf}, Borexino Phase II~\cite{Coloma:2022umy}, Super-K+KamLAND~\cite{Bolanos:2008km}, Liquid Scintillator Neutrino Detector (LSND)~\cite{Davidson:2003ha,LSND:2001akn}, and Taiwan Experiment on Neutrino (TEXONO)~\cite{TEXONO:2010tnr} in the NSI of neutrinos interaction with dense matter through the NC reaction process. It is worth noting that the neutrino scattering data is very sensitive to the NC NSI as reported in Refs.~\cite{Davidson:2003ha,Barranco:2005ps}, which is one of the reasons for focusing on the NC reaction process. Thus, I compute the NDCS and NMFP of the neutrino NSI with the dense matter for various neutrino flavors and different nuclear matter densities. The results of this work will potentially provide information about the NSI NDCS and NMFP in dense matter, which may be useful and relevant to NSI physics in astrophysical phenomena such as neutron stars, supernovae, cosmology, and the early universe. Notably, the NMFP in dense matter is an important input for supernova simulation~\cite{Janka:2012wk}.

This paper is organized as follows. In Sec.~\ref{sec:theory}, I briefly describe the general formalism of the NC NSI for neutrino scattering off free electron gas in dense matter, including the detailed of the NDCS and NMFP constructed from the NSI effective Lagrangian. Section~\ref{sec:num} presents the numerical results of the NSI DCRS and NMFP for various experimental bounds of the vector and axial-vector NSI couplings. Finally, the summary and conclusion are devoted in Sec.~\ref{sec:sum}.

\section{Theoretical Formalism} \label{sec:theory}

\subsection{NSI Lagrangians and cross section}
In this section, I briefly introduce the NSI effective Lagrangian for the elastic $\nu_e-$e scattering in dense matter. Similar to the SM Lagrangian expression for the weak interaction part in Ref.~\cite{Kalempouw-Williams:2005zbp,Horowitz:1990it,Reddy:1998hb}, the NC NSI effective four-fermion Lagrangian for the elastic electron neutrino scattering off the free electron gas reaction process can be written as
\begin{eqnarray}
   \label{eq:nsi1}
    \mathscr{L}_{\rm{NSI}} &=& - \frac{G_F}{\sqrt{2}} \Big[\bar{\nu}_e (k') \gamma^\mu (1-\gamma_5) \nu_e (k) \Big]\Big[ \bar{e}(p') \gamma_\mu \left( \mathcal{E}_{e e} ^{e V} - \mathcal{E}_{e e}^{e A} \gamma_5 \right) e (p) \Big],
\end{eqnarray}
where the subscript of $e$ is the electron neutrino flavors. $G_F = 1.023 \times 10^{-5}/M^2$ is the Fermi constant where $M =$ 939 MeV as the nucleon free mass. The symbols of the $\nu_e (k')$ and $\bar{\nu}_e (k)$ stand for the final and initial neutrino spinors, respectively, and $\bar{e}(p')$ and $e(p)$ are respectively the final and initial lepton spinors, where $k =(k_0, \vec{\mathbf{k}})$ and $p =(p_0, \vec{\mathbf{p}})$ are respectively the initial four-momentum of the neutrinos and electrons, while $k' =(k'_0, \vec{\mathbf{k'}})$ and $p' =(p'_0, \vec{\mathbf{p'}})$ are the final four-momentum of the neutrinos and electrons, respectively.
The parameters $\mathcal{E}_{e e}^{e V} = \mathcal{E}_{e e}^{e L} + \mathcal{E}_{e e}^{e R}$ and $\mathcal{E}_{e e}^{e A} = \mathcal{E}_{e e}^{e L} - \mathcal{E}_{e e}^{e R}$ denote respectively the strength of the NSI vector and axial-vector couplings. In this work, I will concentrate on the NC NSI of the elastic $\nu_e$-e scattering in dense matter.

With the help of the NSI effective Lagrangian in Eq.~(\ref{eq:nsi1}), I derive the neutrino double differential cross section per unit volume $\mathscr{V}$ for NSI of $\nu_e-$e scattering in terms of the electron polarization and lepton tensors (See Appendix for details derivation), and it gives 
\begin{eqnarray}
    \label{eq:nsi-dcrs}
    \frac{1}{\mathscr{V}}\frac{d^3\sigma}{d^2\Omega' dE_\nu'} &=& -\frac{G_F}{32\pi^2} \frac{E_\nu'}{E_\nu} \Im\Big[ L_{\mu \nu} \Pi^{\mu \nu}_{[e]}\Big],
\end{eqnarray}
where $E_\nu$ and $E_\nu'$ stand for the initial and final neutrino energies, respectively. Following the SM lepton tensor of Ref.~\cite{Kalempouw-Williams:2005zbp,Horowitz:1990it,Reddy:1998hb}, the neutrino tensor $L_{\mu \nu}$ for NSI is defined by
\begin{eqnarray}
    \label{eq:lmu1}
    L_{\mu \nu} &=& 8 \left[ 2 k_\mu k_\nu + \left(k\cdot q \right) g_{\mu \nu} - \left( k_\mu q_\nu + q_\mu k_\nu \right) \mp i \epsilon_{\mu \nu \alpha \beta} k^\alpha q^\beta \right],
\end{eqnarray}
where $-$ and $+$ signs in the last term in Eq.~(\ref{eq:lmu1}) are respectively for neutrinos and antineutrinos. $k$ and $q=(q_0,\mathbf{q})$ are four-momentum of the initial neutrinos and transfer, respectively.

In a similar way to the SM polarization tensor for the electrons in dense matter~\cite{Kalempouw-Williams:2005zbp, Horowitz:1990it}, the expression of the electron polarization tensor $\Pi^{\mu \nu}$ for NSI in Eq.~(\ref{eq:lmu1}) can be written as
\begin{eqnarray}
    \label{eq:lmu2}
    \Pi^{\mu \nu}_{[e]} (q) &=&  -i \int \frac{d^4p}{(2\pi)^4} \Tr \Big[ G^{(e)} (p)\gamma^\mu \big(\mathcal{E}_{ee}^{\alpha V} - \mathcal{E}_{ee}{\alpha A} \gamma_5 \big) G^{(e)} (p+q) \gamma^\nu \big( \mathcal{E}_{ee}^{\alpha V} - \mathcal{E}_{ee}^{\alpha A} \gamma_5 \big) \Big],
\end{eqnarray}
where $p=(p_0,\mathbf{p})$ is the initial electron four momentum and $G^{(e)} (p)$ is the electron propagator in matter. The values of the vector and axial-vector NSI couplings for different neutrino flavors obtained from solar, atmospheric, and reactor experiments are summarized in Table~\ref{tab1}. 
The explicit form of the electron propagator at a given electron Fermi momentum $p_F^{(e)}$ is given by
\begin{eqnarray}
    \label{eq:lmu3}
    G^{(e)} (p) &=& \left(p\!\!\!/ + m_e \right) \Big[ \frac{1}{p^{2} - m_e^2 + i\epsilon} + \frac{i\pi}{E_p} \delta \left( p_{0} - E_p \right) \theta \left( p_F^{e} - |\mathbf{p}|\right)\Big],
\end{eqnarray}
where $E_p = \sqrt{\mathbf{p}^2 + m_e^2}$ is the electron energy with $m_e = 0.511$ MeV is the electron free mass. It is worth noting that the free space electron mass is the same as in dense matter.  

After evaluating the contractions of the electron polarization and neutrino tensors of Eq.~(\ref{eq:nsi-dcrs}) by considering the current conservation and translational invariance, the final expression of the NSI NDCS is given by
\begin{eqnarray}
\label{eq:lmu4}
 \frac{1}{\mathscr{V}}\frac{d^3\sigma}{d^2\Omega' dE_\nu'} &=& \frac{G_F^2 E_\nu' q_\mu^2}{4\pi^3 E_\nu} \Big[ A \mathscr{R}_1 + \mathscr{R}_2 + B \mathscr{R}_3 \Big],
\end{eqnarray}
where variables of $A = 2k_0\left( k_0 -q_0\right) + 0.5 q_\mu^2/|\mathbf{q}|^2$ and $B = 2k_0 -q_0$ are kinematic factors and the uncorrelated response functions are respectively defined by
\begin{eqnarray}
    \label{eq:lmu5}
    \mathscr{R}_1 = \mathscr{R}_1^V + \mathscr{R}_1^A &=& \mathcal{E}_{e e}^{{e V} 2} \Big[ \Im \Pi_T^{(e)} + \Im \Pi_L^{(e)} \Big] +  \mathcal{E}_{e e}^{{\alpha A} 2} \Big[ \Im \Pi_T^{(e)} + \Im \Pi_L^{(e)} \Big],\\
    \mathscr{R}_2 = \mathscr{R}_2^V + \mathscr{R}_2^A &=& \mathcal{E}_{e e}^{{e V} 2} \Im \Pi_T^{(e)} + \mathcal{E}_{e e}^{{\alpha A} 2} \Big[ \Im \Pi_T^{(e)} -\Im \Pi_A^{(e)} \Big], \\
    \mathscr{R}_3 &=& \pm 2 \mathcal{E}_{e e}^{\alpha V} \mathcal{E}_{e e}^{{\alpha A}} \Im \Pi_{VA}^{(e)},
\end{eqnarray}
where the signs of $+$ and $-$ in the $\mathscr{R}_3$ are for neutrinos and antineutrinos, respectively. The transverse (T), longitudinal (L), axial (A), and mixed vector-axial (VA) electron polarizations are respectively given by
\begin{eqnarray}
    \label{eq:lmu6}
    \Im \Pi_T^{(e)} &=& \frac{1}{4\pi |\mathbf{q}|} \Bigg[ \left(m_e^2 + \frac{q_\mu^2}{2} + \frac{q_\mu^4}{4|\mathbf{q}|^2} \right) \left( E_F - E^* \right) + q_0 \frac{q_\mu^2}{2|\mathbf{q}|^2} \left( E_F^2 - E^{*2} \right) + \frac{q_\mu^2}{3|\mathbf{q}|^2} \left( E_F^3 - E^{*3} \right)\Bigg],\\
    \Im \Pi_L^{(e)} &=& \frac{q_\mu^2}{2\pi |\mathbf{q}|^3} \Bigg[\frac{q_\mu^2}{4} \left( E_F - E^*\right) + \frac{q_0}{2} \left(E_F^2 -E^{*2} \right) + \frac{1}{3} \left( E_F^3 - E^{*3} \right) \Bigg],\\
    \Im \Pi_A^{(e)} &=& \frac{m_e^2}{2\pi |\mathbf{q}|} \left( E_F - E^* \right), \\
    \Im \Pi_{VA}^{(e)} &=& \frac{q_\mu^2}{8 \pi |\mathbf{q}|^3} \Bigg[ q_0 \left( E_F - E^* \right) + \left( E_F^2 - E^{*2} \right) \Bigg],
\end{eqnarray}
where $E^* = \min \left[ E_F,E_{\rm{max}}\right]$ with $E_F = \sqrt{p_F^{(e) 2} + m_e^2}$ and $E_{\rm{max}} = \max \left[ E_F -q_0, \frac{1}{2} \left(|\mathbf{q}|\sqrt{1- 4m_e^2/q_\mu^2}-q_0 \right) \right]$. Next, with this NDCS formulation, I will numerically compute the NSI NMFP of the process NC $\nu-$e scattering in Sec.~\ref{NSINMFP}.

\subsection{Neutrino mean free path} \label{NSINMFP}
Now, I am in a position to compute the inverse of the NMFP ($\lambda$) by integrating the double NSI NDCS in Eq.~(\ref{eq:lmu4}) over the time and vector components of the neutrino momentum transfer and it gives
\begin{eqnarray}
    \label{eq:lmu7}
    \lambda^{-1} (E_{\nu}) &=& \int^{2E_\nu -q_0}_{q_0} d|\mathbf{q}| \int_0^{2E_\nu} dq_0  \frac{|\mathbf{q}|}{E_\nu' E_\nu} 2\pi \Bigg[ \frac{1}{\mathscr{V}} \frac{d^3\sigma}{d^2\Omega' dE_\nu'} \Bigg],
\end{eqnarray}
where the initial and final neutrino energy are related as $E_\nu' = E_\nu -q_0$. A detailed explanation of the determination of the lower and upper limits of the integral can be found in Refs.~\cite{Hutauruk:2021cgi,Horowitz:1990it}. Note that in the NSI NMFP calculation, we set the $E_\nu = $ 5 and 10 MeV, adapted from Refs.~\cite{Kalempouw-Williams:2005zbp,Horowitz:1990it,Reddy:1998hb,Hutauruk:2023nqc,Hutauruk:2020mhl,Hutauruk:2021cgi}.

\subsection{Experimental bounds on NSI couplings} \label{subNSIbounds}
In this section the stringent constraints on the NSI $\mathcal{E}_{e e}^{e V}$, $\mathcal{E}_{e e}^{e A}$, $\mathcal{E}_{e e}^{e L}$, and $\mathcal{E}_{e e}^{e R}$ couplings obtained from various experiments are presented. The NSI parameters are collected from the different source experiments such as the solar neutrino, atmospheric neutrino, reactor, and long-baseline neutrino experiments. They are the Boroxino~\cite{Agarwalla:2012wf,Coloma:2022umy}, Super-Kamiokande and KamLAND~\cite{Bolanos:2008km}, LSND~\cite{Davidson:2003ha,LSND:2001akn}, and TEXONO~\cite{TEXONO:2010tnr} experiments. The compilation of the bounds of flavor diagonal NC NSI couplings for electron neutrino $\nu_e$ is respectively summarized in Table~\ref{tab1}.

\begin{table}[htb]
  \caption{Experimental bounds for electron neutrino $\nu_e$ NSI couplings with 90\% C.L.}
  \label{tab1}
  \addtolength{\tabcolsep}{10.8pt}
  \begin{tabular}{ccccc} 
    \hline \hline
    Experiments & $\mathcal{E}_{ee}^{eV}$ &  $\mathcal{E}_{ee}^{eA}$ & $\mathcal{E}_{ee}^{eL}$ &  $\mathcal{E}_{ee}^{eR}$ \\[0.2em] 
    \hline
    Boroxino Phase I~\cite{Agarwalla:2012wf}  & X & X & $[-0.046,0.053]$  & $[-0.21,0.16]$\\ 
     Boroxino Phase II~\cite{Coloma:2022umy} & X & X & $[-1.37,-1.29] \oplus [0.03,0.06]$ & $[-0.23,0.07]$\\ 
    Super-K+KamLAND~\cite{Bolanos:2008km} & X & X & $[-0.021, 0.052]$  & $[-0.18,0.51]$  \\
     LSND~\cite{Davidson:2003ha,LSND:2001akn} & X & X & $[-0.07,0.11]$  & $[-1.0,0.5]$  \\
    TEXONO~\cite{TEXONO:2010tnr} & X & X & $[-1.53,0.38]$  & $[-0.07,0.08]$   
    \\ \hline \hline
  \end{tabular}
\end{table}  

Next, all available experimental bounds for the NSI vector and axial-vector couplings in Table~\ref{tab1} will be evaluated in the NDCS and NMFP. It is worth noting that, in this work, I simply choose the lower and upper bounds of the experimental data and their combinations from the same experiment in the computations i.e. the combination of the lower-lower, lower-upper, upper-lower, and upper-upper limits of $\mathcal{E}_{ee}^{eL}$ and $\mathcal{E}_{ee}^{eR}$. The symbol of $X$ indicates that no data for $\mathcal{E}_{ee}^{eV}$ and $\mathcal{E}_{ee}^{eA}$ is directly measured in the experiment. In this work, they are straightforwardly determined from the values of $\mathcal{E}_{ee}^{eV}$ and $\mathcal{E}_{ee}^{eR}$ via the equations $\mathcal{E}_{ee}^{eV} = \mathcal{E}_{ee}^{eL} + \mathcal{E}_{ee}^{eR}$ and $\mathcal{E}_{ee}^{eA} = \mathcal{E}_{ee}^{eL} - \mathcal{E}_{ee}^{eR}$. It is worth noting that, in the calculations of the NDCS and NMFP SM, we use the axial-vector coupling $C_A = g_L -g_R = -1/2$ and vector coupling $C_V = g_L+g_R = -1/2 + 2 \sin^2 \theta_W$, where $\theta_W$ is the weak mixing (Weinberg) angle, and $g_L = -1/2 + \sin^2 \theta_W$ and $g_R = \sin^2 \theta_W$ are the left and right chiral projection couplings in the SM, respectively. Here we used $\sin^2 \theta_W =$ 0.231~\cite{ParticleDataGroup:2022pth}.

\section{Numerical results and Discussions} \label{sec:num}
Here the numerical results for the NSI NDCS and NMFP for the NC $\nu_e-$e scattering with free electron gas in dense matter are presented in Figs.~\ref{fig1}-\ref{fig5} and Figs.~\ref{fig6}-\ref{fig8}, respectively. The NSI NDCS and NMFP of $\nu_e$ are computed by using the initial neutrino energy $E_\nu =$ 5 MeV, which is a typical neutrino energy in neutron star (NS)~\cite{Horowitz:1990it} and fixed values of the three component--momentum transfer $|\mathbf{q}|=$ 2.5 MeV. I also compute the NSI NDCS and NMFP of $\nu_e$ for $E_\nu =$ 10 MeV as used in Ref.~\cite{Horowitz:1990it}.

\subsection{Differential cross section}
\begin{figure}
    \centering
    \includegraphics[width=1\textwidth]{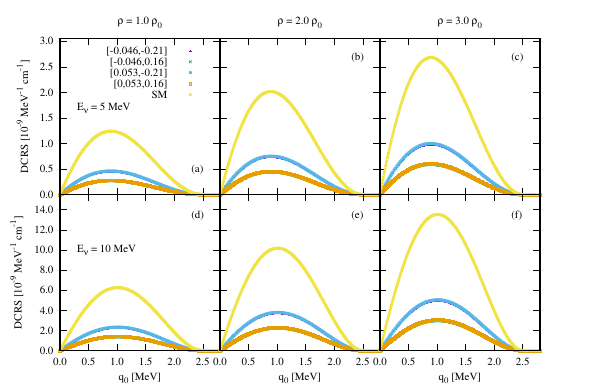}
    \caption{NDCS ($d^3 \sigma/\mathscr{V} d^2\Omega' dq_0$) for different nuclear densities with the initial neutrino energy $E_\nu =$ 5 MeV; (a) $\rho =$ 1.0 $\rho_0$, (b) $\rho =$ 2.0 $\rho_0$, and (c) $\rho =$ 3.0 $\rho_0$ (upper panel), while the lower panel for $E_\nu =$ 10 MeV. Note that the NDCS is computed using the vector and axial-vector NSI couplings extracted from Boroxino Phase I experiment~\cite{Agarwalla:2012wf}. SM represents the NDCS for the Standard Model.}
    \label{fig1}
\end{figure}
The NDCS for testing the NSI couplings constrained from the Boroxino Phase I experiment~\cite{Agarwalla:2012wf} (see Table~\ref{tab1}) is computed. Results for the DCRS for different nuclear densities and initial neutrino energy are shown in Fig.~\ref{fig1}. Figure~\ref{fig1} shows that the NDCS for the NSI couplings [$\mathcal{E}_{ee}^{eL},\mathcal{E}_{ee}^{eR}$] = [$-0.046,-0.21$] (purple point) and [$\mathcal{E}_{ee}^{eL},\mathcal{E}_{ee}^{eR}$] = [$0.053,-0.21$] (blue point) calculated at $E_\nu =$ 5 and 10 MeV, respectively are not significantly different. The reason is that the vector $\mathcal{E}_{ee}^{eV}$ and axial-vector $\mathcal{E}_{ee}^{eA}$ NSI couplings of the two coupling combinations have rather the same values and similar signs, resulting [$\mathcal{E}_{ee}^{eV}, \mathcal{E}_{ee}^{eA}$] = [$-0.256,0.164$], and [$\mathcal{E}_{ee}^{eV}, \mathcal{E}_{ee}^{eA}$] = [$-0.157,0.263$]. Similar features, for both neutrino energies, are shown by the NDCS with the NSI couplings [$\mathcal{E}_{ee}^{eL},\mathcal{E}_{ee}^{eR}$] = [$-0.046,0.16$] (green point) and [$\mathcal{E}_{ee}^{eL},\mathcal{E}_{ee}^{eR}$) = ($0.053,0.16$] (orange point). These coupling combinations give [$\mathcal{E}_{ee}^{eV}, \mathcal{E}_{ee}^{eA}$] = [$0.114,-0.206$], and [$\mathcal{E}_{ee}^{eV}, \mathcal{E}_{ee}^{eA}$] = [$0.213,-0.107$], respectively, which have an opposite sign with the two NSI couplings mentioned earlier. Figure~\ref{fig1}(a)-(f) explicitly shows that the NDCS with $\mathcal{E}_{ee}^{eV} > \mathcal{E}_{ee}^{eA}$ is lower than that with $\mathcal{E}_{ee}^{eV} < \mathcal{E}_{ee}^{eA}$. It indicates that the axial-vector NSI coupling increases the NDCS, while the vector NSI coupling has a role in decreasing the cross-section.

Another interesting feature shown in Figs.~\ref{fig1}(a)-(c) for $E_\nu =$ 5 MeV is the NDCS increases as the nuclear density increases. This trend is followed by the NDCS of $E_\nu =$ 10 MeV as shown in Figs.~\ref{fig1}(d)-(f), but they are only different in magnitude. The NDCS increases by a factor of 5 at the peak as the initial neutrino energy increases by comparing the NDCSs in Figs.~\ref{fig1}(a)-(c) and~\ref{fig1}(d)-(f) for appropriate nuclear densities. Comparing with the SM NDCS (yellow point), it is shown that the NSI NDCS for all NSI couplings is lower than that for the SM. The difference is more pronounced as the nuclear density and neutrino energy increase.

\begin{figure}
    \centering
    \includegraphics[width=1\textwidth]{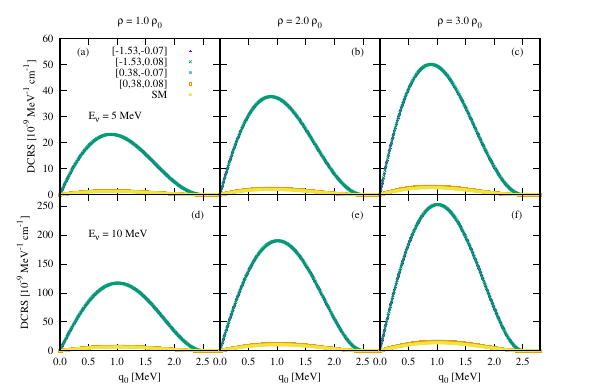}
    \caption{ Same as in Fig.~\ref{fig1} but the NDCS computed using the vector and axial-vector NSI couplings extracted from TEXONO experiment~\cite{TEXONO:2010tnr}.}
    \label{fig2}
\end{figure}

The results of the NDCS with various values of the NSI couplings constrained by the TEXONO experiment~\cite{TEXONO:2010tnr} for different densities and initial neutrino energies are depicted in Figs.~\ref{fig2}(a)-(f). With those NSI couplings from the TEXONO experiment, I obtain the vector and axial vector couplings for [$\mathcal{E}_{ee}^{eL},\mathcal{E}_{ee}^{eR}$] = [$-1.53,-0.07$] (purple point) and [$\mathcal{E}_{ee}^{eL},\mathcal{E}_{ee}^{eR}$] = [$-1.53,0.08$] (green point) are [$\mathcal{E}_{ee}^{eV},\mathcal{E}_{ee}^{eA}$] = [$-1.60,-1.46$] and [$\mathcal{E}_{ee}^{eV},\mathcal{E}_{ee}^{eA}$] = [$-1.46,-1.61$], respectively, where both have the negative values. The other two NSI couplings have positive values of the vector and axial-vector couplings for this respective left and right projection coupling strength [$\mathcal{E}_{ee}^{eL},\mathcal{E}_{ee}^{eR}$] = [$0.38,-0.07$] (blue point) and [$\mathcal{E}_{ee}^{eL},\mathcal{E}_{ee}^{eR}$] = [$0.38,0.08$] (orange point). This results the [$\mathcal{E}_{ee}^{eV},\mathcal{E}_{ee}^{eA}$] = [$0.31,0.45$] and [$\mathcal{E}_{ee}^{eV},\mathcal{E}_{ee}^{eV}$] = [$0.46,0.3$], respectively.

As consequences the NDCS for the [$\mathcal{E}_{ee}^{eL},\mathcal{E}_{ee}^{eR}$] = [$-1.53,-0.07$] and [$\mathcal{E}_{ee}^{eL},\mathcal{E}_{ee}^{eR}$] = [$-1.53,0.08$] is larger than that for [$\mathcal{E}_{ee}^{eL},\mathcal{E}_{ee}^{eR}$] = [$0.38,-0.07$] and [$\mathcal{E}_{ee}^{eL},\mathcal{E}_{ee}^{eR}$] = [$0.38,0.08$] as indicated in Fig.~\ref{fig2}(a)-(f). Again this indicates that the vector and axial-vector NSI couplings compete with each other as previously explained. The competition between the vector and axial-vector NSI couplings with positive values is larger than that with both negative values. This gives rise to the NDCS with the positive values of the vector and axial-vector couplings being smaller as clearly shown in Fig.~\ref{fig2}(a)-(f). 

Figure~\ref{fig2}(a)-(f) shows the NDCS of $E_\nu =$ 5 MeV for different nuclear densities. Similar to the NDCS of the Boroxino Phase I, the NDCS increases when the nuclear density increases. The DCRS also increases when the initial neutrino energy increases. The cross-section increases by a factor of $5$ for the appropriate densities from $E_\nu =$ 5 to 10 MeV. In contrast to the NSI NDCS for the Boroxino Phase I experiment (See Fig.~\ref{fig1}), the NSI NDCS for all couplings extracted from the TEXONO experiment is larger than that for the SM NDCS. This is, indeed, because of their NSI vector and axial-vector coupling difference. The large differences of the NSI to SM NDCSs are given by [$\mathcal{E}_{ee}^{eL},\mathcal{E}_{ee}^{eR}$] = [$-1.53,-0.07$] and [$\mathcal{E}_{ee}^{eL},\mathcal{E}_{ee}^{eR}$] = [$-1.53,0.08$], respectively. 

\begin{figure}
    \centering
    \includegraphics[width=1\textwidth]{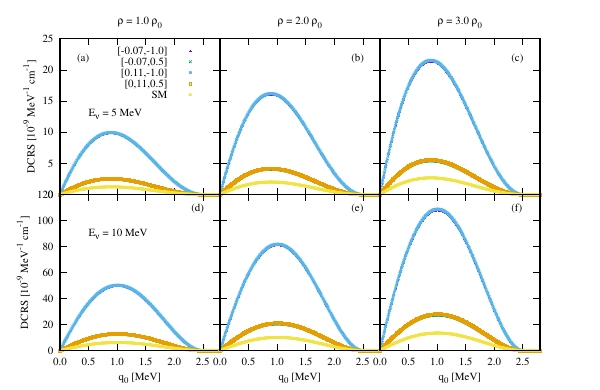}
    \caption{ Same as in Fig.~\ref{fig1} but the NDCS computed using the vector and axial-vector NSI couplings extracted from LSNO experiment~\cite{Davidson:2003ha,LSND:2001akn}.}
    \label{fig3}
\end{figure}

Next, I evaluate the NDCS for the vector and axial-vector NSI couplings extracted from the LSNO experiments~\cite{Davidson:2003ha,LSND:2001akn}. Results for the NDCS for different nuclear densities and initial neutrino energies are given in Fig.~\ref{fig3}(a)-(f). The NDCS for [$\mathcal{E}_{ee}^{eL}, \mathcal{E}_{ee}^{eR}$] = [$-0.07,-1.0$] (purple point) and [$\mathcal{E}_{ee}^{eL}, \mathcal{E}_{ee}^{eR}$] = [$0.11,-1.0$] (blue point) are rather the same. This can be understood because they have similar values of the vector and axial-vector NSI couplings, [$\mathcal{E}_{ee}^{eV}, \mathcal{E}_{ee}^{eA}$] = [$-1.07,0.93$] and [$\mathcal{E}_{ee}^{eV}, \mathcal{E}_{ee}^{eA}$] = [$-0.89,1.11$], respectively as found in Fig.~\ref{fig1} with the Boroxino Phase I, where the vector and axial-vector couplings have negative and positive signs, respectively. The NDCS for [$\mathcal{E}_{ee}^{eL}, \mathcal{E}_{ee}^{eR}$] = [$-0.07,0.5$] (green point) and [$\mathcal{E}_{ee}^{eL}, \mathcal{E}_{ee}^{eR}$] = [$0.11,0.5$] (orange point) also look the same, giving [$\mathcal{E}_{ee}^{eV}, \mathcal{E}_{ee}^{eA}$] = [$0.43,-0.57$] and [$\mathcal{E}_{ee}^{eV}, \mathcal{E}_{ee}^{eA}$] = [$0.61,-0.39$]. As expected the NDCS for [$\mathcal{E}_{ee}^{eL}, \mathcal{E}_{ee}^{eR}$] = [$-0.07,0.5$] and [$\mathcal{E}_{ee}^{eV}, \mathcal{E}_{ee}^{eA}$] = [$0.11,0.5$] is relatively smaller than that for [$\mathcal{E}_{ee}^{eL}, \mathcal{E}_{ee}^{eR}$] = [$-0.07,-1.0$] and [$\mathcal{E}_{ee}^{eL}, \mathcal{E}_{ee}^{eR}$] = [$0.11,-1.0$]. Again, it shows similar features as the previous result that the NDCS for both $E_\nu =$ 5 and 10 MeV increases as the nuclear density increases. Also, the NDCS increases by a factor of $5$ for suitable nuclear densities when the initial neutrino energy increases. In comparison with the NDCS of the SM (yellow point), it is shown that the NDCS for all NSI coupling of the LSNO experiment is larger than that found in the SM.

\begin{figure}
    \centering
    \includegraphics[width=1\textwidth]{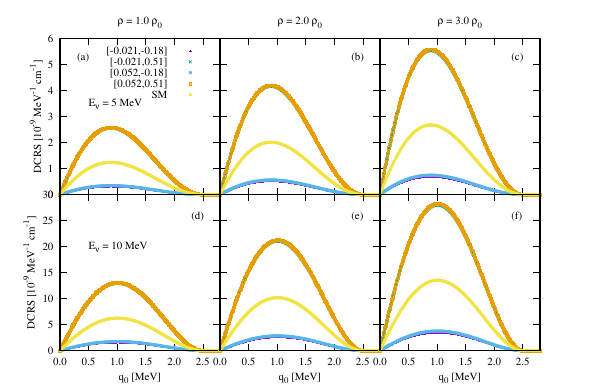}
    \caption{ Same as in Fig.~\ref{fig1} but the NDCS computed using the vector and axial-vector NSI couplings extracted from SuperKamLAND experiment~\cite{Bolanos:2008km}.}
    \label{fig4}
\end{figure}

Further testing for the vector and axial factor extracted from the SuperKamLAND (superK $+$ KamLAND) experiment~\cite{Bolanos:2008km} to NDCS is given in Fig.~\ref{fig4}. As the Boroxino Phase I and LSNO, the typical vector and axial-vector NSI couplings obtained from the combinations of the $\mathcal{E}_{ee}^{eL}$ and $\mathcal{E}_{ee}^{eR}$ have similar features, but different in magnitude. The values of vector and axial-vector NSI coupling for [$\mathcal{E}_{ee}^{eL}, \mathcal{E}_{ee}^{eR}$] = [$-0.021,-0.18$], [$\mathcal{E}_{ee}^{eL}, \mathcal{E}_{ee}^{eR}$] = [$-0.021,0.51$], [$\mathcal{E}_{ee}^{eL}, \mathcal{E}_{ee}^{eR}$] = [$0.052,-0.18$], and [$\mathcal{E}_{ee}^{eL}, \mathcal{E}_{ee}^{eR}$] = [$0.052,0.51$] are respectively [$\mathcal{E}_{ee}^{eV}, \mathcal{E}_{ee}^{eA}$] = [$-0.201,0.159$], [$\mathcal{E}_{ee}^{eV}, \mathcal{E}_{ee}^{eA}$] = [$0.489,-0.531$], [$\mathcal{E}_{ee}^{eV}, \mathcal{E}_{ee}^{eA}$] = [$-0.128,0.232$], and [$\mathcal{E}_{ee}^{eV}, \mathcal{E}_{ee}^{eA}$] = [$0.562,-0.458$]. 

In Fig.~\ref{fig4}(a), it is shown the NDCS for [$\mathcal{E}_{ee}^{eL}, \mathcal{E}_{ee}^{eR}$] = [$-0.021,0.51$] and [$\mathcal{E}_{ee}^{eL}, \mathcal{E}_{ee}^{eR}$] = [$0.052,0.51$] is relatively the same. Similarly, it is also shown by the NDCS of [$\mathcal{E}_{ee}^{eL}, \mathcal{E}_{ee}^{eR}$] = [$-0.021,-0.18$] and [$\mathcal{E}_{ee}^{eL}, \mathcal{E}_{ee}^{eR}$] = [$0.052,-0.18$]. Figure~\ref{fig4}(a) also shows that the NDCS of [$\mathcal{E}_{ee}^{eL}, \mathcal{E}_{ee}^{eR}$] = [$-0.021,0.51$] and [$\mathcal{E}_{ee}^{eL}, \mathcal{E}_{ee}^{eR}$] = [$0.052,0.51$] is larger than that of [$\mathcal{E}_{ee}^{eL}, \mathcal{E}_{ee}^{eR}$] = [$-0.021,-0.18$] and [$\mathcal{E}_{ee}^{eL}, \mathcal{E}_{ee}^{eR}$] = [$0.052,-0.18$]. Again, the NDCS increases as nuclear density increases and neutrino energy increases, as shown in Figs.~\ref{fig4}(a)-(c) and Figs.\ref{fig4}(a) and (d), respectively.

Comparing with the SM NDCS (yellow point) it is shown that the NDCS of [$\mathcal{E}_{ee}^{eL}, \mathcal{E}_{ee}^{eR}$] = [$-0.021,0.51$] and [$\mathcal{E}_{ee}^{eL}, \mathcal{E}_{ee}^{eR}$] = [$0.052,0.51$] is larger than that of the SM, while the NDCS of [$\mathcal{E}_{ee}^{eL}, \mathcal{E}_{ee}^{eR}$] = [$-0.021,-0.18$] and [$\mathcal{E}_{ee}^{eL}, \mathcal{E}_{ee}^{eR}$] = [$0.052,-0.18$] is relatively lower than that of the SM.

\begin{figure}
    \centering
    \includegraphics[width=1\textwidth]{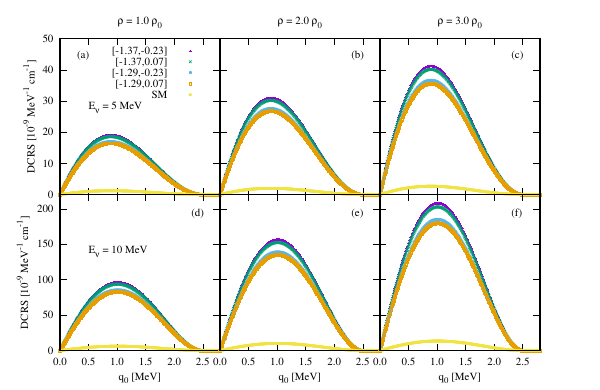}\\
     \includegraphics[width=1\textwidth]{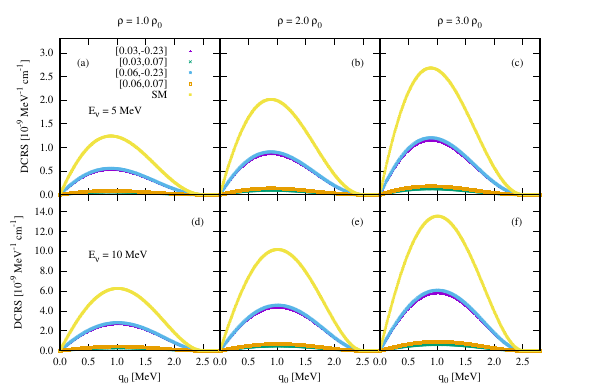} 
    \caption{ Same as in Fig.~\ref{fig1} but the NDCS computed using the vector and axial-vector NSI couplings extracted from Boroxino Phase II experiment~\cite{Coloma:2022umy}.}
    \label{fig5}
\end{figure}

The NDCS results for the vector and axial-vector NSI coupling extracted from the Boroxino Phase II experiment~\cite{Coloma:2022umy} shown in the upper panel of Fig.~\ref{fig5}(a)-(f) and lower panel of Fig~\ref{fig5}(a)-(f). The NSI coupling combinations are taken from the two regions of the couplings as shown in Table~\ref{tab1}, where the $\mathcal{E}_{ee}^{eL}$ has two regions. The first region is $\mathcal{E}_{ee}^{eL} = [-1.37,-1.29]$ and the second region lies on $\mathcal{E}_{ee}^{eL} = [0.03,0.06]$. The NDCS for the combinations of the first region of $\mathcal{E}_{ee}^{eL}= [-1.37,-1.29]$ and $\mathcal{E}_{ee}^{eR}$ is depicted in the upper panel of Fig.~\ref{fig5}(a)-(f) while that for the combinations of the second region of $\mathcal{E}_{ee}^{eL}= [0.03,0.06]$ and $\mathcal{E}_{ee}^{eR}$ is given in the lower panel of Fig.~\ref{fig5}(a)-(f). 

The left and right chiral projection couplings of [$-1.37,-0.23$], [$-1.37,0.07$], [$-1.29,-0.23$], and [$-1.29,0.07$] yield the vector and axial-vector couplings with values [$-1.6,-1.14$], [$-1.3,-1.44$], [$-1.52,-1.06$], and [$-1.22,-1.36$], respectively. In the upper panel Fig.~\ref{fig5}(a)-(f) it is shown that the NDCS of the vector and axial-vector NSI couplings [$\mathcal{E}_{ee}^{eL}, \mathcal{E}_{ee}^{eR}$]= [$-1.37,-0.23$] (purple point) and [$\mathcal{E}_{ee}^{eL}, \mathcal{E}_{ee}^{eR}$]= [$-1.37,0.07$] (green point) is a bit larger than that for [$\mathcal{E}_{ee}^{eL}, \mathcal{E}_{ee}^{eR}$]= [$-1.29,-0.23$] (blue point) and [$\mathcal{E}_{ee}^{eL}, \mathcal{E}_{ee}^{eR}$]= [$-1.29,0.07$] (orange point). The features of the vector and axial-vector NSI couplings are the same as those of the TEXONO data, both have negative vector and axial-vector NSI couplings. The difference between  [$\mathcal{E}_{ee}^{eL}, \mathcal{E}_{ee}^{eR}$]= [$-1.37,-0.23$] and [$\mathcal{E}_{ee}^{eL}, \mathcal{E}_{ee}^{eR}$]= [$-1.37,0.07$] become rather visible as the nuclear density increases as shown in Figs.~\ref{fig5}(c) and ~\ref{fig5}(f). This behavior is followed by the vector and axial-vector NSI couplings with respective values [$\mathcal{E}_{ee}^{eL}, \mathcal{E}_{ee}^{eR}$]= [$-1.29,-0.23$] and [$\mathcal{E}_{ee}^{eL}, \mathcal{E}_{ee}^{eR}$]= [$-1.29,0.07$]. As a previous result of the NDCS, here it is also found that the NDCS increases as the nuclear density and initial neutrino energy increase. Additionally, the interesting result is shown by the NDCSs of all couplings extracted from the Boroxino Phase II (first region, see Table~\ref{tab1}), which are larger than that found in the SM (yellow point).

The left and right chiral projection bounds extracted from the second region of the Boroxino Phase II produce the vector and axial-vector couplings with values [$\mathcal{E}_{ee}^{eV}, \mathcal{E}_{ee}^{eA}$]= [$-0.2,0.26$], [$\mathcal{E}_{ee}^{eV}, \mathcal{E}_{ee}^{eA}$]= [$0.1,-0.04$], [$\mathcal{E}_{ee}^{eV}, \mathcal{E}_{ee}^{eA}$]= [$-0.17,0.29$], and [$\mathcal{E}_{ee}^{eV}, \mathcal{E}_{ee}^{eA}$]= [$0.13,-0.01$]. The NDCS results of the NSI couplings are shown in the lower panel of Fig.~\ref{fig5}. It shows that the NDCS for [$\mathcal{E}_{ee}^{eL}, \mathcal{E}_{ee}^{eR}$]= [$0.03,-0.23$] (purple point) and [$\mathcal{E}_{ee}^{eL}, \mathcal{E}_{ee}^{eR}$]= [$0.06,-0.23$] (blue point) is larger than that of [$\mathcal{E}_{ee}^{eL}, \mathcal{E}_{ee}^{eR}$]= [$0.03,0.07$] (green point) and [$\mathcal{E}_{ee}^{eL}, \mathcal{E}_{ee}^{eR}$]= [$0.06,0.07$] (orange point). The NDCS difference between [$\mathcal{E}_{ee}^{eL}, \mathcal{E}_{ee}^{eR}$]= [$0.03,-0.23$] and [$\mathcal{E}_{ee}^{eL}, \mathcal{E}_{ee}^{eR}$]= [$0.06,-0.23$] increases as the nuclear density increases as shown in the lower panel of Figs.~\ref{fig5}(a)-(c). This feature is also shown in the NDCS of the [$\mathcal{E}_{ee}^{eL}, \mathcal{E}_{ee}^{eR}$]= [$0.03,0.07$] and [$\mathcal{E}_{ee}^{eV}, \mathcal{E}_{ee}^{eA}$]= [$0.06,0.07$]. Also, the NDCS increases as the initial neutrino energy increases as previously found. The NDCS for all NSI couplings from the Boroxino Phase II (second region, see Table~\ref{tab1}) is lower than that for the SM (yellow point).

At this point, one concludes that the NDCS increases as the nuclear density and initial neutrino energy increase. The NDCS strongly depends on the vector and axial-vector NSI couplings that are constrained by various recent experiments, producing different NDCS. In other words, the NDCS is sensitive to the vector and axial-vector NSI couplings. More measurements of the vector and axial-vector couplings are necessary to have a rigorous prediction on the neutrino scattering off free gas electrons in dense matter or other astrophysical objects. I expect that the NDCS in the dense matter will be relatively the same for different experiments if they have precise values of the couplings.

\subsection{Neutrino mean free path}
Now I turn to present the numerical results of the NSI NMFP. Results of the NSI NMFP as a function of the nuclear density for different vector and axial-vector NSI couplings and initial neutrino energy are given in Fig.~\ref{fig6}-\ref{fig8}. It is worth noting that the NMFP is computed numerically from the NDCS through Eq.~(\ref{eq:lmu7}).

In Fig.~\ref{fig6}(a)-(b), it is shown that the NMFP of the vector and axial-vector coupling extracted from the Boroxino Phase I for $E_\nu =$ 5 and 10 MeV, respectively. As the NDCS results for Boroxino Phase I, the NMFP for the [$\mathcal{E}_{ee}^{eR}, \mathcal{E}_{ee}^{eL}$]=[$-0.046,-0.21$] (purple point) and [$\mathcal{E}_{ee}^{eL}, \mathcal{E}_{ee}^{eR}$]=[$0.053,-0.21$] (blue point) are rather the same, while the NMFP for [$\mathcal{E}_{ee}^{eL}, \mathcal{E}_{ee}^{eR}$]=[$-0.046,0.16$] (green point) is the same as [$\mathcal{E}_{ee}^{eL}, \mathcal{E}_{ee}^{eR}$]=[$0.053,0.16$] (orange point). The NMFP for [$\mathcal{E}_{ee}^{eL}, \mathcal{E}_{ee}^{eR}$]=[$0.053,-0.21$] and [$\mathcal{E}_{ee}^{eL}, \mathcal{E}_{ee}^{eR}$]=[$-0.046,-0.21$] are rather smaller than that for [$\mathcal{E}_{ee}^{eL}, \mathcal{E}_{ee}^{eR}$]=[$-0.046,0.16$] and [$\mathcal{E}_{ee}^{eL}, \mathcal{E}_{ee}^{eR}$]=[$0.053,0.16$]. It is also shown that the NMFP for all couplings decreases as the nuclear density increases. The NMFP ($\lambda$) with $E_\nu =$ 5 MeV for [$\mathcal{E}_{ee}^{eL}, \mathcal{E}_{ee}^{eR}$]=[$0.053,-0.21$] and [$\mathcal{E}_{ee}^{eV}, \mathcal{E}_{ee}^{eA}$]=[$-0.046,-0.21$] is around 900 km at $\rho =$ 3.0 $\rho_0$ and $\lambda \simeq$ 1200 km for [$\mathcal{E}_{ee}^{eL}, \mathcal{E}_{ee}^{eR}$]=[$-0.046,0.16$] and [$\mathcal{E}_{ee}^{eL}, \mathcal{E}_{ee}^{eR}$]=[$0.053,0.16$]. With this $\lambda \simeq $ 900 and 1200 km, in the NSI scenario context, it is expected the neutrino will easily escape from the neutron star, where i.e. in general neutron stars have $R_{\rm{NS}} =$ 12.39$_{-0.98}^{+1.30}${km~\cite{Riley:2021pdl} (from astrophysical observation). The NMFP decreases as the initial neutrino energy increases as shown in Fig.~\ref{fig6}(b) as expected. The interesting result is shown by the SM NMFP, which is relatively the same as the NMFP of [$-0.046,-0.21$] and [$0.053,-0.21$] at $\rho \gtrsim 0.5 \rho_0$, while the NMFP for [$-0.046,0.16$] and [$0.053,016$] is larger than that of the SM at $\rho \gtrsim 0.5 \rho_0$.
\begin{figure}
    \centering
    \includegraphics[width=1\textwidth]{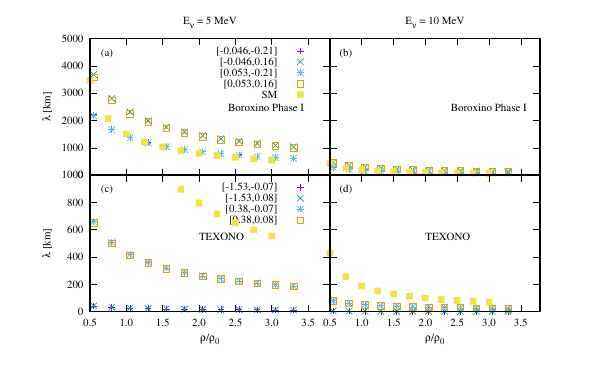}
    \caption{NMFP as a function of nuclear densities for $E_\nu =$ 5 and 10 MeV. The NMFP is computed using the NSI couplings from the Boroxino Phase I (upper panel) and TEXONO~\cite{TEXONO:2010tnr} (lower panel) experiments. SM is the NMFP result for the Standard Model.}
    \label{fig6}
\end{figure}

The NMFP for the vector and axial-vector NSI couplings extracted from the TEXONO experiment~\cite{TEXONO:2010tnr} are shown in Figs.~\ref{fig6}(c)-(d). Similar to the NMFP for the Boroxino Phase I, the NMFP decreases as the nuclear density increases, but they are different in magnitude. The NMFP for the vector and axial vector couplings [$\mathcal{E}_{ee}^{eL},\mathcal{E}_{ee}^{eR}$]=[$0.38,0.08$] and [$\mathcal{E}_{ee}^{eL},\mathcal{E}_{ee}^{eR}$]=[$0.38,-0.07$] are relatively the same, while the NMFP of [$\mathcal{E}_{ee}^{eL},\mathcal{E}_{ee}^{eR}$]=[$-1.53,-0.07$] is somewhat the same as [$\mathcal{E}_{ee}^{eV},\mathcal{E}_{ee}^{eA}$]=[$-1.53,0.08$]. Again, the NMFP decreases as the nuclear density and initial neutrino energy increase as in Fig.~\ref{fig6}(d). The NMFP for [$0.38,-0.07$] and [$0.38,0.08$] is larger than that for [$-1.53,-0.07$] and [$-1.53,0.08$]. Figures~\ref{fig6}(c) and (d) show that the SM NDCS is larger than that for all couplings extracted from the TEXONO experiment~\cite{TEXONO:2010tnr}.

Next the NMFP for the vector and axial-vector couplings obtained from the LSNO and SuperKamLAND experiments as a function of the nuclear density for different initial neutrino energy is given in Fig.~\ref{fig7}. In general, the NMFP result has a similar trend as other NMFP for the couplings of different experiments, where the NMFP decreases as the nuclear density and initial energy of neutrino increase. Of course, different vector and axial vector coupling will give different NMFP predictions as shown in Figs.~\ref{fig7}(a)-(d) as previously found in the NMFP results. For the LSNO experiment, the NDCS of all couplings is lower than that found in the SM (yellow point) as in Figs.~\ref{fig7}(a) and (b). The NDCS for [$-0.07,0.5$] and [$0.11,0.5$] is slightly larger than that for [$-0.07,-1.0$] and [$0.11,-1.0$], where the NDCS for [$-0.07,0.5$] and [$0.11,0.5$] are relatively the same, followed by the NDCS for [$-0.07,-1.0$] and [$0.11,-1.0$].

Finally, the NMFP results for the vector and axial-vector NSI couplings from Boroxino phase II are given in Fig.~\ref{fig8}. Interesting behavior is shown in Fig.~\ref{fig8}(a), giving the NMFP for all vector and axial-vector couplings is relatively the same. This can be understood because the NDCS for all NSI coupling is relatively the same as shown in Fig.~\ref{fig5}(a)-(f). It is worth noting that increasing the NDCS leads to decreasing the NMFP. Figure~\ref{fig8} shows the NMFP for [$-1.37,-0.23$] (purple point) and [$-1.37,0.07$] (green point) are relatively the same, while the NDCS of [$-1.29,-0.23$] and [$-1.29,0.07$] has relatively similar values. In Figs.\ref{fig8}(a) and (b) it is also shown that the NMFP for the SM is larger than that for all couplings extracted from Boroxino Phase II (first bounds of $\mathcal{E}_{ee}^{eL}$).

For the Boroxino Phase II (second bounds of $\mathcal{E}_{ee}^{eL}$) the NMFP for the left and right chiral projection couplings of [$\mathcal{E}_{ee}^{eL}, \mathcal{E}_{ee}^{eR}$] = [$0.03,0.07$] and [$\mathcal{E}_{ee}^{eL}, \mathcal{E}_{ee}^{eR}$] = [$0.06,0.07$] are larger than that for [$0.03,-0.23$] and [$0.06,-0.23$], but the largest is given by [$0.03,0.07$] as shown in Fig.~\ref{fig8}(c). The NDCS for [$0.03,-0.23$] and [$0.06,0.07$] are relatively the same. Other things are still similar to other NMFPs it decreases as the nuclear density and initial energy increase. I also found that the SM NMFP is lower than that for [$[0.03,0.07$] and [$0.06,0.07$], but slightly larger than [$0.03,-0.23$] and [$0.06,-0.23$].

\begin{figure}
    \centering
    \includegraphics[width=1\textwidth]{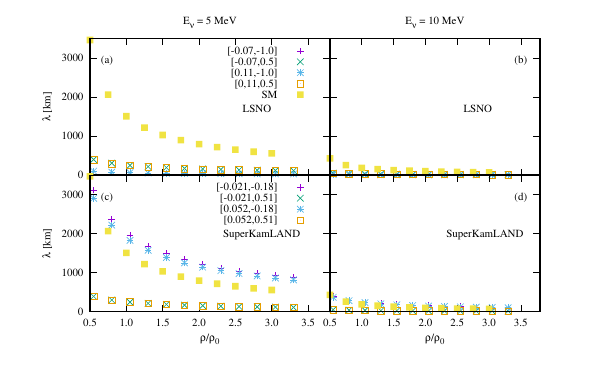}
    \caption{Same as in Fig.~\ref{fig6}, but for the LSNO~\cite{Davidson:2003ha,LSND:2001akn} (upper panel) and SuperKamLAND~\cite{Bolanos:2008km} (lower panel) experiments.}
    \label{fig7}
\end{figure}

\begin{figure}
    \centering
    \includegraphics[width=1\textwidth]{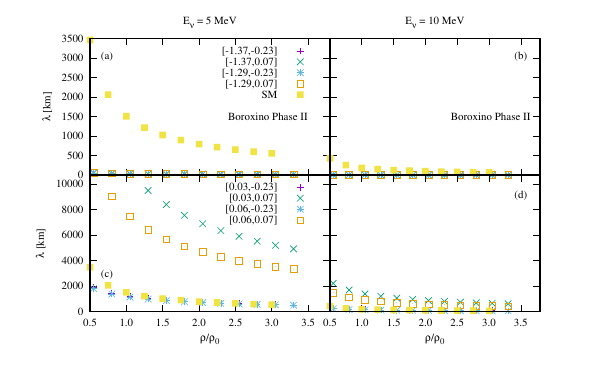}
    \caption{Same as in Fig.~\ref{fig6}, but for the Boroxino Phase II experiment~\cite{Coloma:2022umy}.}
    \label{fig8}
\end{figure}

\section{Summary and future perspectives} \label{sec:sum}
In this paper, I have explored the effects of the vector and axial-vector NSI couplings extracted from various recent experiments in the NSI neutrino scattering off the free electron gas in dense matter. I compute the NSI NDCS and NMFP in dense matter for various vector and axial-vector NSI couplings. It is worth noting that the vector and axial-vector NSI couplings from different experiments are unique, meaning the values of the NSI couplings are almost different for various experiments, which may lead to new physics.

For the Boroxino Phase I~\cite{Agarwalla:2012wf}, I found the NCDS of all NSI couplings is lower than that of the SM. In contrast with the Boroxino Phase I, for the NSI coupling extracted from the TEXONO experiment~\cite{TEXONO:2010tnr}, the results of the NDCS are larger than that for the SM. The large differences to the SM are given by [$-1.53,-0.07$] and [$-1.53,0.008$]. Similar to the TEXONO experiment, the NDCS for all the couplings extracted from the LSNO experiment~\cite{Davidson:2003ha,LSND:2001akn} is relatively larger than that found in the SM, but different in magnitude. For the SuperKamLAND experiment~\cite{Bolanos:2008km}, comparing with the SM NDCS I found that the NDCS for [$-0.021,0.51$] and [$0.052,0.51$] are larger than that of the SM, while the NDCS for [$-0.21,-0.18$] and [$0.052,-0.18$] is lower than that for the SM. I also found that the NDCSs for the Boroxino Phase II~\cite{Coloma:2022umy} (first region of $\mathcal{E}_{ee}^{eL}$) are larger than that of the SM, while the NDCS for the Boroxino Phase II~\cite{Coloma:2022umy} (second region of $\mathcal{E}_{ee}^{eL}$) are smaller than that of the SM. In most cases, I found that the NDCS increases as the nuclear density and neutrino energy increase. Increasing the DCRS leads to a decrease in the NMFP. 

In the NMFP results, for the Boroxino Phase I, I found the NMFPs for [$-0.046,0.16$] and [$0.053,0.16$] are larger than that found in the SM, while the NMFPs for [$-0.046,0.21$] and [$0.053,-0.21$] are relatively the same as that for the SM at $\rho \gtrsim$ 0.5 $\rho_0$. For the TEXONO experiment, the NMFPs for all NSI couplings are lower than that for the SM. Similarly, this behavior of the NMFP is found in the LSNO experiment. 

For the SuperKamLAND experiment, the NMFP of [$-0.021,0.18$] and [$0.052,0.18$] is found to be larger than that of SM, while for [$-0.021,0.51$] and [$0.052,0.51$] the NMFP is lower than that for the SM. I also found the NMFP for the Boroxino Phase II (first region of $\mathcal{E}_{ee}^{eL}$) is lower than the NMFP for the SM. In the second region of $\mathcal{E}_{ee}^{eL}$, it is found that the NMFP of [$0.03,0.07$] and [$0.06,0.07$] is larger than that of the SM, whereas for [$0.03,-0.23$] and [$0.06,-0.23$] it is found that the NMFP is relative the same as that of the SM. Overall the NMFP decreases as the nuclear density and neutrino energy increase as expected.

The results of the present work provide useful information for the neutrino-NSI interaction in neutron stars, supernovas, and other astrophysical objects as well as supernova simulation~\cite{Janka:2012wk}. It was expected that the neutrino oscillation dominated by matter so the result of this study is also very relevant to such a topic. Additionally, the prediction of this current work on the NSI NMFP and NDCS for different NSI couplings in dense matter gives an intuition of the NSI NMFP's appearance in dense matter. In the future, I will extend the study to neutrino oscillation in matter through the neutrino interaction. The related works are still in progress and will appear elsewhere.

\section*{ACKNOWLEDGEMENT}
This work was supported by the National Research Foundation of Korea (NRF) grants funded by the Korean government (MSIT) (2018R1A5A1025563, 2022R1A2C1003964, and 2022K2A9A1A0609176). I thank Anto Sulaksono (Universitas Indonesia) for valuable discussions and stimulating ideas.

\appendix
\section{Electron Polarization Tensor}
The electron propagator in the polarization tensor in Eq.(\ref{eq:lmu2}) can be also written as follows:
\begin{eqnarray}
    \label{eq:appen1}
    G^{(e)} (p) = G_F^{(e)} (p) + G_D^{(e)} (p) = \frac{p\!\!\!/+m_e}{p^2 -m_e^2 + i\epsilon} + \frac{i\pi}{E_p} \delta (p_0 -E_p) \theta (p_F^e -|p|) (p\!\!\!/ +m_e),
\end{eqnarray}
where $G_F^{(e)} (p)$, and $G_D^{(e)} (p)$ are the usual Feynmann and density-dependent of the target electron propagators, respectively. Substituting this medium propagator of the electron target in Eq.~(\ref{eq:appen1}), which is equivalent with Eq.~(\ref{eq:lmu3}), into the electron polarization tensor in Eq.~(\ref{eq:lmu2}), one has
\begin{eqnarray}
\label{eq:appen2}
\Pi^{\mu \nu}_{[e]} (q) &=& -i \int \frac{d^4q}{(2\pi)^4} \mathrm{Tr}\Big[ \big[G_{F}(p) + G_D (p) \big] J^\mu \big[ G_F (p+q) + G_D (p+q) \big] J^\nu \Big], \nonumber \\
&=& -i \int \frac{d^4q}{(2\pi)^4} \mathrm{Tr} \Big[ G_F (p) J^\mu G_F (p+q) J^\nu + G_F (p) J^\mu G_D (p+q) J^\nu \nonumber \\
&+& G_D (p) J^\mu G_F (p+q) J^\nu + G_D (p) J^\mu G_D (p+q) J^\nu \Big],
\end{eqnarray}
where $J^\mu = \gamma^\mu \big( \mathcal{E}_{ee}^{eV} - \mathcal{E}_{ee}^{eA} \big) \gamma_5$. It is worth noting that in this work, we use the mean-field approach, so the divergent term of $G_F (p) J^\mu G_F (p+q) J^\nu$ can be ignored. The electron polarization can be written as
\begin{eqnarray}
    \label{eq:append3}
    \Pi^{\mu \nu}_{[e]} (q) &=& -i \int \frac{d^4q}{(2\pi)^4} \mathrm{Tr} \Big[ G_F (p) J^\mu G_D (p+q) J^\nu + G_D (p) J^\mu G_F (p+q) J^\nu \nonumber \\
    &+& G_D (p) J^\mu G_D (p+q) J^\nu \Big],  \nonumber \\
    &=& -i \int \frac{d^4q}{(2\pi)^4} \mathrm{Tr} \Big[ \frac{1}{2} G_D (p) J^\mu G_D (p+q) J^\nu + G_F (p) J^\mu G_D (p+q) J^\nu \Big] \nonumber \\
    &-& i \int \frac{d^4q}{(2\pi)^4} \mathrm{Tr} \Big[ \frac{1}{2} G_D (p) J^\mu G_D (p+q) J^\nu + G_D (p) J^\mu G_F (p+q) J^\nu \Big].
\end{eqnarray}

After plugging the vertices of $J^{\mu}$ and the medium electron propagator into Eq.~(\ref{eq:appen2}), 
\begin{eqnarray}
    \label{eq:appen3a}
    \Pi^{\mu \nu}_{[e]} (q) &=& -i \int \frac{d^4q}{(2\pi)^4} \mathrm{Tr} \Big[ g_D (p) \big( p\!\!\!/ + m_e \big) \big( \gamma^\mu \mathcal{E}_{ee}^{eV} - \mathcal{E}_{ee}^{eA} \gamma^\mu \gamma_5 \big) g_F (p+q) \big( p\!\!\!/ + q\!\!\!/ + m_e \big) \big( \gamma^\nu \mathcal{E}_{ee}^{eV} - \mathcal{E}_{ee}^{eA} \gamma^\nu \gamma_5 \big)  \nonumber \\
    &+& \frac{1}{2} g_D (p) \big( p\!\!\!/ + m_e \big) \big( \gamma^\mu \mathcal{E}_{ee}^{eV} - \mathcal{E}_{ee}^{eA} \gamma^\mu \gamma_5 \big) g_D (p+q) \big( p\!\!\!/ + q\!\!\!/ + m_e \big) \big( \gamma^\nu \mathcal{E}_{ee}^{eV} - \mathcal{E}_{ee}^{eA} \gamma^\nu \gamma_5 \big) \Big] + \Big[ p \rightarrow -p\Big], \nonumber \\
    &=& -i \int \frac{d^4q}{(2\pi)^4} \mathrm{Tr} \Big[ g_D (p) g_F (p+q) + \frac{1}{2} g_D (p) g_D (p+q) \Big] \nonumber \\
    &\times& \Big[ \big( p\!\!\!/ + m_e \big) \big( \mathcal{E}_{ee}^{eV} \gamma^\mu - \mathcal{E}_{ee}^{eA} \gamma^\mu \gamma_5 \big) \big( p\!\!\!/ q\!\!\!/ + m_e \big) \big( \mathcal{E}_{ee}^{eV} \gamma^\nu - \mathcal{E}_{ee}^{eA} \gamma^\nu \gamma_5 \big) \Big] + \Big[ p \rightarrow - p \Big], \nonumber \\
    &=& -i \int \frac{d^4q}{(2\pi)^4} \Big[ g_D (p) g_F (p+q) + \frac{1}{2} g_D (p) g_D (p+q) \Big] \mathcal{F}^{\mu \nu} (p,p+q) + \Big[ p \rightarrow -p \Big].
\end{eqnarray}

Now, we compute a trace in Eq.~(\ref{eq:appen3a}) and trace calculations of Eq.~(\ref{eq:appen3a}) are given as
\begin{eqnarray}
    \label{eq:appen4}
\mathcal{F}^{\mu \nu} (p,p+q) &=& \mathrm{Tr} \Big[ \big(p\!\!\!/ + m_e \big) \big(\mathcal{E}_{ee}^{eV} \gamma^\mu - \mathcal{E}_{ee}^{eA} \gamma^\mu \gamma_5 \big) \big( p\!\!\!/ + q\!\!\!/ + m_e \big) \big( \mathcal{E}_{ee}^{eV} \gamma^\nu - \mathcal{E}_{ee}^{eA} \gamma^\nu \gamma_5\big) \Big], \nonumber \\
&=& \big(\mathcal{E}_{ee}^{eV} \big)^2 \mathcal{F}^{\mu \nu}_{eV} (p,p+q) - 2 \mathcal{E}_{ee}^{eA} \mathcal{E}_{ee}^{eV} \mathcal{F}_{eV-eA}^{\mu \nu} (p,p+q) + \big(\mathcal{E}_{ee}^{eA} \big)^2 \mathcal{F}^{\mu \nu}_{eA} (p,p+q), 
\end{eqnarray}
where the vector, axial, and axial vector are respectively defined as
\begin{eqnarray}
    \label{eq:appen5}
    \mathcal{F}^{\mu \nu}_{eV} (p,p+q) &=& 4 \Big[ 2 p^\mu p^\nu + p^\mu q^\nu + p^\nu q^\mu - p\cdot q g^{\mu \nu} \Big], \\
    \mathcal{F}^{\mu \nu}_{eA} (p,p+q)  &=& 4 \Big[ 2 p^\mu p^\nu + p^\mu q^\nu + p^\nu q^\mu - p \cdot q g^{\mu \nu} - 2 m_e^2 g^{\mu \nu} \Big],\\
    \mathcal{F}^{\mu \nu}_{eV-eA} (p,p+q) &=& -4i \epsilon^{\mu \nu \rho \sigma} p_\rho q_\sigma.
\end{eqnarray}

Similarly, we compute the second term of Eq.~(\ref{eq:append3}) after we apply the shifting variable $\Big[ p \rightarrow -p \Big]$ to $F^{\mu \nu} (p,p+q)$ an done has
\begin{eqnarray}
\label{eq:appen6}
 -i \int \frac{d^4q}{(2\pi)^4} \mathrm{Tr} \Big[ G_D (p) J^\mu G_F (p-q) J^\nu + \frac{1}{2} G_D (p) J^\mu G_D (p-q) J^\nu \Big],
\end{eqnarray}
we then substitute $G_D (p)$, $G_F(p-q)$, and $G_D (p-q)$, it gives
\begin{eqnarray}
    \label{eq:append7}
    &=& -i \int \frac{d^4q}{(2\pi)^4} \mathrm{Tr} \Big[ g_D (p) \big( p\!\!\!/ + m_e \big) \big( \gamma^\mu \mathcal{E}_{ee}^{eV} - \mathcal{E}_{ee}^{eA} \gamma^\mu \gamma_5 \big) g_F (p-q) \big( p\!\!\!/ -q\!\!\!/ + m_e \big) \big( \gamma^\nu \mathcal{E}_{ee}^{eV} - \mathcal{E}_{ee}^{eA} \gamma^\nu \gamma_5 \big)  \nonumber \\
    &+& \frac{1}{2} g_D (p) \big( p\!\!\!/ + m_e \big) \big( \gamma^\mu \mathcal{E}_{ee}^{eV} - \mathcal{E}_{ee}^{eA} \gamma^\mu \gamma_5 \big) g_D (p-q) \big( p\!\!\!/ -q \!\!\!/ + m_e \big) \big( \mathcal{E}_{ee}^{eV} \gamma^\nu - \mathcal{E}_{ee}^{eA} \gamma^\nu \gamma_5 \big) \Big], \nonumber \\
    &=& -i \int \frac{d^4q}{(2\pi)^4} \Big[g_D (p) g_F (p-q) + \frac{1}{2} g_D (p) g_D (p-q) \Big] \mathcal{F}^{\mu \nu} (p,p-q),
\end{eqnarray}
where the trace of $\mathcal{F}^{\mu \nu} (p,p-q)$ is given as follows
\begin{eqnarray}
    \label{eq:appen8}
    \mathcal{F}^{\mu \nu} (p,p-q) &=& \mathrm{Tr} \Big[ \big(p\!\!\!/ +m_e \big) \big( \gamma^\mu \mathcal{E}_{ee}^{eV} - \mathcal{E}_{ee}^{eA} \gamma^\mu \gamma_5 \big) \big( p\!\!\!/ -q\!\!\!/ + m_e \big) \big( \gamma^\nu \mathcal{E}_{ee}^{eV} - \mathcal{E}_{ee}^{eA} \gamma^\nu \gamma_5 \big)\Big], \nonumber \\
    &=& \big( \mathcal{E}_{ee}^{eV} \big)^2 \mathcal{F}^{\mu \nu}_{eV} (p,p-q) -2 \mathcal{E}_{ee}^{eV} \mathcal{E}_{ee}^{eA} \mathcal{F}^{\mu \nu}_{eV-eA} (p,p-q) + \big( \mathcal{E}_{ee}^{eA} \big)^2 \mathcal{F}^{\mu \nu}_{eA} (p,p-q),
\end{eqnarray}
where $\mathcal{F}^{\mu \nu}_{eV} (p,p-q)$, $\mathcal{F}^{\mu \nu}_{eA} (p,p-q)$, and $\mathcal{F}^{\mu \nu}_{eV-eA} (p,p-q)$ are respectively defined as
\begin{eqnarray}
    \mathcal{F}^{\mu \nu}_{eV} (p,p-q) &=& 4 \Big[ 2 p^\mu p^\nu - p^\mu q^\nu -p^\nu q^\mu + p \cdot q g^{\mu \nu} \Big],\\
    \mathcal{F}^{\mu \nu}_{eA} (p,p-q) &=& 4 \Big[ 2 p^\mu p^\nu - p^\mu q^\nu - p^\nu q^\mu + p \cdot q g^{\mu \nu} - 2 m_e g^{\mu \nu} \Big], \\
    \mathcal{F}^{\mu \nu}_{eV-eA} (p,p-q) &=& 4 i \epsilon^{\rho \mu \sigma \nu } \Big[ p_\rho p_\sigma - p_\rho q_\sigma \Big]
\end{eqnarray}

We turn into the calculation of the electron propagator of Eqs.~(\ref{eq:appen3a}) and~(\ref{eq:append7}). The contractions of the propagators are compactly given as 
\begin{eqnarray}
\label{eq:appen9}
    \Pi^{\mu \nu}_{e} (q) &=& -i \int \frac{d^4p}{(2\pi)^4} \Big[ g_D (p) g_F (p+q) + \frac{1}{2} g_D (p) g_D (p+q) \Big] \mathcal{F}^{\mu \nu} (p, p+q) + \Big[ p \rightarrow -p \Big],
\end{eqnarray}
where the contractions of the propagators in Eq.~(\ref{eq:appen9}) are respectively given as
\begin{eqnarray}
\label{eq:appen10}
g_D (p) g_F (p \pm q) &=& \frac{i\pi}{E_p} \delta \big( p_0 - E_p \big) \theta \big( p_F^{e} - |p| \big) \delta \big( p_0 \pm q_0 - E_{p\pm q} \big) \Big[ \frac{\mathcal{P}}{\big( p\pm q\big) -m_e^2} - \frac{i\pi}{2E_{p \pm q}} \Big], \\
g_D (p) g_D (p \pm q) &=& - \frac{\pi^2}{E_p E_{p\pm q}} \delta \big( p_0 - E_p \big) \theta \big(p_F^{e} -|p| \big) \delta \big( p_0 \pm q_0 E_{p \pm q} \big) \theta \big( p_F - |p \pm q| \big).
\end{eqnarray}
where the $\mathcal{P}$ is the \textit{Principle value}. After substituting the contractions of the propagators of Eq.~(\ref{eq:appen10}) into Eq.~(\ref{eq:appen9}) and a long calculation for the real and imaginer part of the electron polarization, we obtain 
\begin{eqnarray}
    \label{eq:appen11}
    \Pi^{\mu \nu}_{e} (q) &=& \big(\mathcal{E}_{ee}^{eV}\big)^2 \Pi^{\mu \nu}_{IeV} (q) + \big(\mathcal{E}_{ee}^{eA} \big)^2 \Pi^{\mu \nu}_{IeA} (q) - 2 \mathcal{E}_{ee}^{eV} \mathcal{E}_{ee}^{eA} \Pi^{\mu \nu}_{IeV-IeA} (q),
\end{eqnarray}
where
\begin{eqnarray}
    \Pi^{\mu \nu}_{IeV} (q) &=& \frac{-i\pi^2}{2 E_p E_{p+q}} \int \frac{d^4q}{(2\pi)^4} \mathcal{F}^{\mu \nu}_{eV} (p,p+q) \delta \big( p_0 - E_p \big) \theta \big( p_F^e - |p| \big) \theta \big( |p+q| - p_F^e \big) \delta \big( p_0 +q_0 - E_{p+q} \big), \\
    \Pi^{\mu \nu}_{IeA} (q) &=& \frac{-i\pi^2}{2 E_p E_{p+q}} \int \frac{d^4q}{(2\pi)^4} \mathcal{F}^{\mu \nu}_{eA} (p,p+q) \delta \big( p_0 - E_p \big) \theta \big( p_F^e - |p| \big) \theta \big( |p+q| - p_F^e \big) \delta \big( p_0 +q_0 - E_{p+q} \big), \\
     \Pi^{\mu \nu}_{IeV-IeA} (q) &=& \frac{-i\pi^2}{2 E_p E_{p+q}} \int \frac{d^4q}{(2\pi)^4} \mathcal{F}^{\mu \nu}_{eV-eA} (p,p+q) \delta \big( p_0 - E_p \big) \theta \big( p_F^e - |p| \big) \theta \big( |p+q| - p_F^e \big) \delta \big( p_0 +q_0 - E_{p+q} \big).
\end{eqnarray}

In the calculation, we choose the frame defined as $q^\mu = \big( q^0, |q|,0,0 \big)$, and $p^\mu =\big(p^0 = E, p^x, p^y, p^z\big)$, where $p^x = |p| \cos \theta$, $p^y = |p| \sin \theta \cos \phi$, and $p^z = |p| \sin \theta \sin \phi$. With this frame, then the components of $\mathcal{F}_{eV}^{\mu \nu} (p,p +q)$ can be defined as
\begin{eqnarray}
    \mathcal{F}^{\mu \nu}_{eV} (p,p+q) &=& 4 \Big[ 2 p^\mu p\nu + p^\mu q^\nu + p^\nu q^\mu - p\cdot q g^{\mu \nu} \Big], \\
    \mathcal{F}^{00}_{eV} (p,p+q) &=& 4 \Big[ 2E^2 + Eq_0 + |p||q| \cos \theta\Big], \\
    \mathcal{F}^{11}_{eV} (p,p+q) &=& 4 \Big[ 2|p|^2 \cos^2 \theta + |p||q| \cos \theta + Eq_0 \Big], \\
    \mathcal{F}^{22}_{eV} (p,p+q) &=& 4 \Big[ 2 |p|^2 \sin^2 \theta \cos^2 \phi + Eq_0 + |p||q| \cos \theta \Big], \\
    \mathcal{F}^{33}_{eV} (p,p+q) &=& 4 \Big[ 2|p|^2 \sin^2 \theta \sin^2 \phi + Eq_0 + |p||q| \cos \theta \Big].
\end{eqnarray}

By substituting these components into $\Pi^{\mu \nu}_{IeV} (q)$, the imaginer polarization insertion can be written in the matrix form
\begin{eqnarray}
\Pi^{\mu \nu}_{IeV} = \begin{bmatrix}
    \Pi^{00}_{IeV} & \Pi^{01}_{IeV} & 0 & 0 \\
     \Pi^{10}_{IeV} & \Pi^{11}_{IeV} & 0 & 0 \\
     0 & 0 & \Pi^{22}_{IeV} & 0 \\
     0 & 0 & 0 & \Pi^{33}_{IeV} 
\end{bmatrix}.
\end{eqnarray}
 It is worth noting that these electron propagators satisfy the current conservation $q_\mu \Pi^{\mu \nu}_{IeV} (q) =$ 0. With this, the vector electron polarization can be decomposed into electron transversal and longitudinal polarizations, which are respectively defined as 
 \begin{eqnarray}
     \Pi_T^{(e)} = \Pi^{22} = \Pi^{33} &=& \frac{1}{4\pi |\mathbf{q}|} \Big[ \big( m_e^2 + \frac{q_\mu^2}{2} + \frac{q_\mu^4}{4 |\mathbf{q}|} \big) \big( E_F - E^* \big) + \frac{q_0 q_\mu^2 }{2|\mathbf{q}|^2} \big( E_F^2 -E^{*2} \big) + \frac{q_\mu^2}{3 |\mathbf{q}|^2} \big( E_F^3 - E^{*3} \big) \Big], \\
     \Pi_L^{(e)} = - \frac{q_\mu^2}{|q|^2} \Pi^{00} &=& \frac{q_\mu^2}{2\pi |\mathbf{q}|^3} \Big[ \frac{q_\mu^2}{4} \big( E_F - E^*\big) + \frac{q_0}{2} \big( E_F^2 - E^{*2} \big) + \frac{1}{3} \big( E_F^3 - E^{*3} \big) \Big] 
 \end{eqnarray}

Analogous to the vector polarization of electrons, the axial polarization of electrons can be simply defined in terms of the longitudinal and transversal polarizations,
\begin{eqnarray}
    \Pi^{L}_{IeA} (q) &=& \Pi^{L}_{IeV} (q) + \Pi^{A} (q), \\
    \Pi^{T}_{IeA} (q) &=& \Pi^{T}_{IeV} (q) - \Pi_A (q),
\end{eqnarray}
 where $\Pi_A (q)$ is defined as
 \begin{eqnarray}
     \Pi_A (q) &=& \frac{m_e^2}{2\pi |\mathbf{q}|} \Big[ E_F - E_F^*\Big].
 \end{eqnarray}
 and shortly, the axial-vector polarization of electron $\Pi^{\mu \nu}_{IeV-IeA}$ is defined as 
 \begin{eqnarray}
     \Pi^{\mu \nu}_{IeV-IeA} (q) &=& \frac{q_\mu^2}{8\pi |\mathbf{q}|^3} \Big[ \big( E_F^2 - E^{*2} \big) + q_0 \big( E_F - E^* \big) \Big].
 \end{eqnarray}
To this point, the total imaginer polarization of electrons can be written in terms of the vector, axial, and axial-vector polarization as follows:
\begin{eqnarray}
    \Pi^{\mu \nu}_{e} (q) &=& \big(\mathcal{E}_{ee}^{eV} \big)^2 \Pi^{\mu \nu}_{IeV} (q) + \big( \mathcal{E}_{ee}^{eA} \big)^2 \Pi^{\mu \nu}_{IeA} (q) - 2 \mathcal{E}_{ee}^{eV} \mathcal{E}_{ee}^{eA} \Pi^{\mu \nu}_{IeV-IeA} (q)
\end{eqnarray}

\section{Contractions}
 Here I present the detailed contractions between the lepton and the medium electron polarization tensors of Eq.~(\ref{eq:nsi-dcrs}). The contraction is given as
 \begin{eqnarray}
     L{\mu \nu} \Pi^{\mu \nu}_{e} = \mathrm{Im} \Big[ L_{\mu \nu} \Pi^{\mu \nu}_{e} \Big] &=& \big(\mathcal{E}_{ee}^{eA} \big)^2 L_{\mu \nu} \Pi^{\mu \nu}_{IeA} + \big( \mathcal{E}_{ee}^{eV} \big)^2 L_{\mu \nu} \Pi^{\mu \nu}_{IeV} -2 \mathcal{E}_{ee}^{eV} \mathcal{E}_{ee}^{eA} L_{\mu \nu} \Pi^{\mu \nu}_{IeV-IeA}, \\
     &=& -8 q_\mu^2 \Big[ A \mathscr{R}_1 + \mathscr{R}_2 + B \mathscr{R}_3],
 \end{eqnarray}
where $\mathscr{R}_1$, $\mathscr{R}_2$, and $\mathscr{R}_3$ is the same definition as in Eq.~(\ref{eq:lmu5}).


\end{document}